\documentstyle[12pt]{article}
\input epsf.sty
 1

\catcode`\@=11 
\makeatletter
\def\@seccntformat#1{\csname the#1\endcsname.\hskip 1em}

\makeatother
\begin{document}

\thispagestyle{empty}
\begin{flushright}

{\footnotesize\renewcommand{\baselinestretch}{.75}
           SLAC-PUB-8444 \\
           April 2000 \\
}
\end{flushright}

\vspace {0.5cm}

\begin{center}
{\large \bf Measurements of $Z^0$ Electroweak Couplings at SLD$^*$}

\vspace {2.0cm}

 {\bf Hermann St\"angle}

\vspace {0.2cm}

University of Massachusetts, Amherst, MA 01003

\vspace {0.6cm}

 {\bf Representing The SLD Collaboration$^{**}$}

\vspace {0.2cm}

Stanford Linear Accelerator Center \\
Stanford University, Stanford, CA~94309 \\

\vspace{2.0cm}
{\bf Abstract}
\end{center}

\renewcommand{\baselinestretch}{1.2}

We present a summary of the results of several electroweak
measurements performed by the SLD experiment at the Stanford Linear 
Collider (SLC). Most of these results are preliminary 
and are based, unless otherwise indicated, on the full 1993-8 data
set of approximately 550,000 hadronic decays of $Z^0$ bosons 
produced with an average electron beam polarization of 73\%.

\vspace{0.7cm}
\begin{center}
{\it 
Invited talk presented at the 
14th Rencontres de Physique de la Vallee d'Aoste: \\
Results and Perspectives in Particle Physics \\
La Thuile, Italy \\
February 27--March 4, 2000
}
\end{center}

\vfil

\noindent
$^*$Work supported in part by Department of Energy contracts 
DE-FG03-93ER40788 and DE-AC03-76SF00515.
\eject

\section{Introduction}

In the $SU(2)_L \times U(1)$ Standard Model (SM), the
vertex factor for the weak neutral current interaction 
in the $Z^0 \rightarrow f\bar{f}$ process is given by:
\begin{equation}
   -i\frac{g}{\cos{\theta_W}} \frac{\gamma^{\mu}}{2} (v_f - a_f \gamma^5)
\end{equation}
where $v_f$ and $a_f$ denote the vector and axial-vector couplings:
\begin{equation}
 v_f = (c_L^f + c_R^f) = T_f^3 - 2 Q_f \sin^2{\theta_W}, \;\;\;\;\;\;\;\;
 a_f = (c_L^f - c_R^f) = T_f^3. 
\end{equation}
Here $T_f^3$ denotes the $3^{rd}$ component of the fermion weak isospin,
$Q_f$ is the electric charge of the fermion, and $\theta_W$ represents the 
electroweak mixing angle.

From combinations of these couplings the $Z^0$ pole observables 
$A_f$ and $R_f$ can be formed.
$A_f$ represents the extent of parity violation in the 
coupling of the $Z^0$ boson to the fermion of type $f$:
\begin{equation}
   A_f =  \frac{2 v_f a_f}{v_f^2 + a_f^2}
       =  \frac{(c_L^f)^2 -(c_R^f)^2}{(c_L^f)^2 +(c_R^f)^2},
\end{equation}
and $R_f$ denotes the fraction of $Z^0 \rightarrow f\bar{f}$ 
events in hadronic $Z^0$ decays:
\begin{equation}
 R_f = \frac{\Gamma(Z^0 \rightarrow f\bar{f})}
            {\Gamma(Z^0 \rightarrow \mbox{hadrons})} 
      \propto (c_L^f)^2 +(c_R^f)^2.  
\end{equation}
The precise experimental determination of $A_f$ and $R_f$ should 
ultimately verify whether or not the couplings for all 
generations and weak isospin states are described by the theory 
isospin assignments and a universal value of $\sin^2{\theta_W}$.

At Born level, the differential production cross section for
$e^+ e^- \rightarrow Z^0 \rightarrow f\bar{f}$ and longitudinally
polarized electrons can be written as:
\begin{equation}
     \sigma^f(x) =     \frac{d\sigma^f}{dx}
            \propto (1-A_e P_e)(1 + x^2) + 2A_f(A_e-P_e)x
\end{equation}
where $x$ denotes the cosine of the polar angle of the outgoing
fermion $f$ with respect to the incident electron beam direction,
$P_e$ represents the longitudinal polarization of the electron beam,
and the positron beam is assumed unpolarized.
If one measures the polar angle distribution for a given final state $
f\bar{f}$, one can derive the forward-backward production asymmetry:
\begin{equation}
 A^f_{FB}(x) = \frac{\sigma^f(x) - \sigma^f(-x)}{\sigma^f(x) + \sigma^f(-x)}
             = 2 A_f \frac{A_e - P_e}{1 - A_e P_e} \frac{x}{1+x^2}
\end{equation}
which depends on both the initial and final state coupling parameters as
well as on the beam polarization.  For zero polarization, one measures
the product of couplings $A_e A_f$. 
If one measures the distributions in equal luminosity samples taken with
negative $(L)$ and positive $(R)$ beam polarization of magnitude $P_e$,
then one can derive the left-right-forward-backward asymmetry:
\begin{equation}
\tilde{A}^f_{FB}(x) = \frac{(\sigma^f_L(x) + \sigma^f_R(-x)) -
                            (\sigma^f_R(x) + \sigma^f_L(-x))}
                           {(\sigma^f_L(x) + \sigma^f_R(-x)) +
                            (\sigma^f_R(x) + \sigma^f_L(-x))}
                    = 2 |P_e| A_f \frac{x}{1+x^2}
\end{equation}
for which the dependence on the initial state coupling disappears, allowing
a direct measurement of the final state coupling parameter $A_f$.
Thus the presence of electron beam polarization permits 
unique $A_f$ measurements\cite{LEPSLDewwg}, not only
independent of those inferred from the unpolarized forward-backward
asymmetry\cite{LEPSLDewwg}  
which measures the combination $A_e A_f$, but also with a 
statistical advantage of $(P_e/A_e)^2 \sim 25$. 
The initial state coupling is determined most precisely
via the left-right cross section asymmetry:
\begin{equation}
  A_{LR} = \frac{1}{P_e}  
           \frac{\sigma_L - \sigma_R}{\sigma_L + \sigma_R} = A_e
\end{equation}
which yields a very precise measurement of the electroweak mixing 
angle due to $\delta A_e \sim 8\delta\sin^2{\theta_W}$.
The measurement and comparison of $A_f$ for the different charged
lepton species also provides a direct test of lepton universality.

In addition, precise measurements of $A_f$ and $R_f$ can probe the
effect of radiative corrections to the $Z^0$ propagator or the
$Z^0 \rightarrow f\bar{f}$ vertex. The radiative corrections
depend on the masses of top and Higgs, and precise electroweak measurements
can constrain these quantities.
The coupling of the $Z^0$ boson to the $b$ quark
is particularly interesting. Physics beyond the SM may couple more
strongly to 3$^{rd}$ generation fermions, producing larger deviations
in $b$ quark couplings than in other quark couplings.
Since $(c_L^b)^2 \sim 30 (c_R^b)^2 $, $R_b$ has large sensitivity
to possible deviations from the predicted left-handed coupling
of the $Z^0$ boson to the $b$ quark, complementary to $A_b$ which
has greater sensitivity to the right-handed coupling.

\section{Unique Features of SLD/SLC}

The performance of the SLC in the 1997-8 SLD data run has been
excellent, with peak luminosities of $3 \times 10^{30}$~cm$^{-2}$~$s^{-1}$  
(i.e. 20,000 $Z^0$ decays/week). Thus approximately 550,000 $Z^0$
decays have been collected during the 1993-8 data runs.

A general description of the SLD detector can be found in 
Ref.~\cite{SLD}. Here we list several of the unique features which
allow the SLD experiment to perform many competitive electroweak
and heavy flavor measurements:
\begin{itemize}
\item
A highly longitudinally polarized (average $\sim$73\%) electron beam. 
\item
A small and stable beam spot 
(1.5~$\mu$m $\times$ 0.8~$\mu$m $\times$ 700~$\mu$m)
and a high precision 3D~CCD-based pixel vertex detector~\cite{VXD}
allow the interaction point to be determined to 
6~$\mu$m $\times$ 6~$\mu$m $\times$ 25~$\mu$m 
with an impact parameter resolution of 11~$\mu$m $\times$ 23~$\mu$m
($r\phi \times rz$) for high momentum tracks.
\item
Good particle identification provided by the Cherenkov 
Ring Imaging Detector (CRID)~\cite{CRID}.
\end{itemize}

\section{Lepton Coupling Measurements}

\subsection{Left-right Cross Section Asymmetry ($A_{LR}$)}

$A_{LR}$ provides a direct measurement of the initial state
electron coupling, independent of the final state coupling.
No efficiency or acceptance corrections are needed.
The final state identification is relatively unsophisticated,
and since practically all of the data can be used,
$A_{LR}$ can be measured~\cite{JBrau} with high statistical precision.
In fact, due to the high precision on the polarization 
measurement, the result is still statistically limited
($\sim$1.3\% statistical error compared to $\sim$0.65\% systematic
error). The precision on $A_{LR}$ demands extensive cross 
checks to confirm the measurement, and three of the significant
more recent checks have been the secondary, independent
confirmations of the electron polarization measurement,
the verification of the center-of-mass collision energy, 
and the measurement confirming the absence of positron polarization. 
The $A_{LR}$ measurement uses all hadronic events and
is effectively a counting experiment; 
$A_{LR} = \frac{1}{P_e} \frac{N_L - N_R}{N_L + N_R}$ where $N_L$($N_R$)
denotes the number of $Z^0$ decays recorded with left (right)
electron beam polarization.
After applying additional corrections
due to $\gamma$ exchange and $\gamma Z^0$ interference,
we measure $A_{LR}^0 = 0.15108 \pm 0.00218$ and 
$\sin^2{\theta_W^{eff}} = 0.23101 \pm 0.00028$
with the 1992-8 data sample.
The $A_{LR}$ measurement provides the most precise single 
measurement of $\sin^2{\theta_W^{eff}}$ presently available.

\subsection{$A_{lepton}$ from $\tilde{A}^f_{FB}$}

The lepton couplings $A_e$, $A_{\mu}$ and $A_{\tau}$ are measured~\cite{JBrau}
at SLD from leptonic decays of $Z^0$ bosons making use of the
corresponding left-right-forward-backward asymmetry, $\tilde{A}^f_{FB}$, 
for each lepton type. $A_e$ and $A_l$ (with $l = \mu, \tau$) are
extracted simultaneously using a maximum likelihood fit. We obtain 
$A_e = 0.1558 \pm 0.0064$, $A_{\mu} = 0.137 \pm 0.016$,
and $A_{\tau} = 0.142 \pm 0.016$. These results are consistent with 
lepton universality and can be combined to yield
$A_{e\mu\tau} = 0.1523 \pm 0.0057$ corresponding to 
$\sin^2{\theta_W^{eff}} = 0.23085 \pm 0.00073$.

\section {$R_b$ and $R_c$ Measurements}

\begin{figure}[t]
 \centering
 \epsfxsize10cm
 \leavevmode
 \epsffile{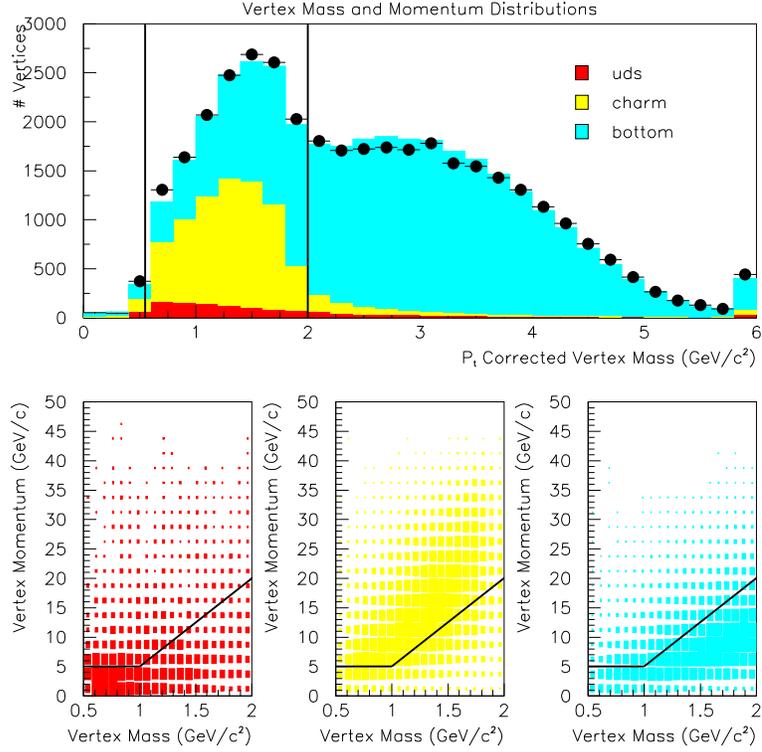}
 \caption {\label{LT_MvtxPvtx}
          Distributions of (top) $M_{vtx}$ with dots representing the data
          and (bottom) $P_{vtx}$ vs. $M_{vtx}$ for $uds$ (left) $c$ (middle)
          and $b$ (right) hemispheres.}
 \end{figure}

The $R_c$~\cite{Rc} and $R_b$~\cite{Rb} measurements 
heavily exploit the excellent
vertexing capabilities of the SLD via a robust and efficient 
topological vertex algorithm~\cite{zvtop}. 
Following a standard hadronic event selection, each event is divided 
into two hemispheres where secondary (and tertiary) vertices are found.
After calculating the $p_t$ corrected vertex invariant mass, $M_{vtx}$, 
hemispheres are tagged as containing a $b$ quark if $M_{vtx} > 2$~GeV/c$^2$
and the secondary vertex is at least 5$\sigma$ away from the 
primary vertex. The hemisphere purity for $b$ events is 
$\sim$98\% with $\sim$50\% efficiency. Events are selected and tagged 
as $b\bar{b}$ if at least one of their hemispheres satisfies these conditions.
Similarly, an event is tagged as $c\bar{c}$ if 
there is at least one track with 3D impact parameter more than 
3$\sigma$ from the primary vertex, 
$0.55$~GeV/c$^2 < M_{vtx} < 2$~GeV/c$^2$,
$P_{vtx} > 5$~GeV/c and $P_{vtx} > 15 M_{vtx}c - 10$~GeV/c.
The hemisphere purity for $c$ events is 
$\sim$70\% with $\sim$16\% efficiency.
Figure~\ref{LT_MvtxPvtx} shows the $M_{vtx}$ and
$P_{vtx}$ vs. $M_{vtx}$ distributions for $uds$, $c$ and $b$ hemispheres.
The charm and bottom tagging efficiencies are self-calibrated
directly from the experimental data using the single, double and mixed 
tag rates. The Monte Carlo is used as input for the $c$ and $uds$
efficiencies in the $b$ tag region. Therefore it is important
to have high purity in the $b$ tagged event sample in order to
reduce the systematic uncertainties due to the modeling
of charm production and decay in the simulation. Hemisphere 
correlations are also derived from the simulation.
We measure $R_b = 0.2159 \pm 0.0014(stat.) \pm 0.0014(syst.)$
and $R_c = 0.1685 \pm 0.0047(stat.) \pm 0.0043(syst.)$.
Approximately 150,000 hadronic $Z^0$ decays from the last part of
the 1998 run have not yet been included in the $R_b$ result.

\section{$A_q$ Measurements}
SLD provides several methods to measure the quark couplings $A_q$,
and their statistical and systematic correlations are taken into account
in the combinations of their results.
All these measurements construct the left-right-forward-backward asymmetries
defined previously and use, in different ways, the secondary vertex
information. The four different techniques employed to measure $A_b$
make use of the jet charge, the kaon charge, the vertex charge, and
the lepton charge to tag the $b$ hemisphere.
To determine the $c$ hemisphere, the four $A_c$ measurements 
employ the kaon or vertex charge, the lepton charge, the soft
pion charge or $D^{(*)}$ meson decays reconstructed exclusively.
$A_s$ is measured at SLD using identified kaons.

\subsection{$A_b$ with Kaon Tag}
The $b \rightarrow c \rightarrow s$ decay chain is exploited in this
measurement~\cite{Ab_Ktag} to tag the sign of the initial $b$ quark.
This measurement heavily relies on the good $K^{\pm}$ 
identification provided by the CRID and the excellent separation
between $K^{\pm}$ coming from the secondary vertex and those from the 
IP. The analyzing power for $b$ events is calibrated from the data.
We obtain $A_b = 0.960 \pm 0.040(stat.) \pm 0.069(syst.)$.

\subsection{$A_b$ with Vertex Charge} 
In this measurement~\cite{Ab_vtxcharge}, 
the sum of the charges of tracks attached to the
reconstructed vertex is used to tag the initial $b$ quark sign.
The analyzing power for $b$ events is calibrated from the data. We measure 
$A_b = 0.897 \pm 0.027(stat.) \pm 0.034(syst.)$
with the 1996-8 data sample. 

\subsection{$A_b$ with Momentum-Weighted Jet Charge}
\begin{figure}[ht]
 \centering
 \epsfxsize7cm
 \leavevmode
 \epsffile{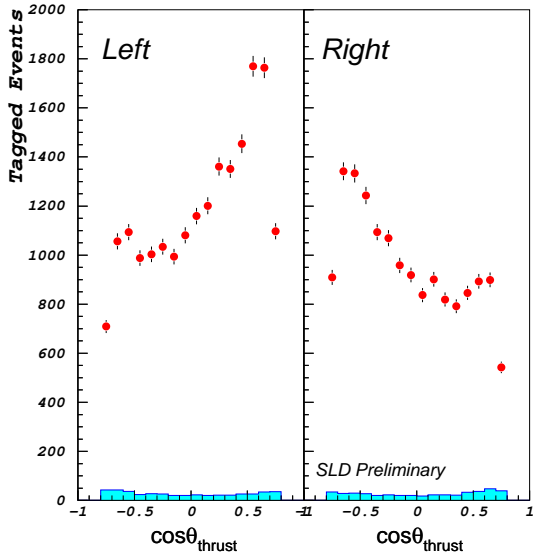}
 \caption {\label{LT_Ab_jetch978}
           Polar angle distributions of the signed thrust axis
           for left-handed and right-handed electron beams for the $A_b$
           with jet charge analysis. Dots represent the data, and 
           the estimated background is indicated by the shaded histograms.}
 \end{figure}

The momentum weighted jet charge~\cite{Ab_mwjetcharge} is defined as:
\begin{equation}
   Q_{diff} = Q_b - Q_{\bar b} = 
            - \sum_{tracks} q_i \cdot \mbox{sgn}
              (\vec{p}_i \cdot \hat{T}) |(\vec{p}_i \cdot \hat{T})|^{\kappa}
\end{equation}
where $\vec{p}_i$ and $q_i$ denote the $i^{th}$ track momentum and charge,
respectively, and $\hat{T}$ represents the direction of the thrust
axis. The coefficient $\kappa$ was chosen to be 0.5 in order to 
maximize the analyzing power of the tag. $Q_{diff}$ is the difference
between the momentum-weighted charges in the two hemispheres.
The analyzing power for $b$ events is calibrated from the data.
The hemisphere correlation is taken from the simulation.
Figure~\ref{LT_Ab_jetch978} shows the polar angle distributions 
of the signed thrust axis for left-handed and right-handed electron beams. 
We measure $A_b = 0.882 \pm 0.020(stat.) \pm 0.029(syst.)$.

\subsection{$A_b$ and $A_c$ with Lepton Tag}
$A_b$ and $A_c$ can be measured by tagging bottom and charm hadrons using
their semileptonic decays~\cite{Acb_leptag}.
The lepton total and transverse momenta with respect to the nearest jet
are employed to calculate the probabilities that the lepton comes from each
of the possible physics processes 
$Z^0 \rightarrow b\bar{b}, b \rightarrow l$;
$Z^0 \rightarrow b\bar{b}, \bar{b} \rightarrow \bar{c} \rightarrow l$; 
$Z^0 \rightarrow b\bar{b}, b \rightarrow \bar{c} \rightarrow l$;
$Z^0 \rightarrow c\bar{c}, \bar{c} \rightarrow l$; 
and background (leptons from light hadron decays, photon conversions,
misidentified hadrons). The lepton charge provides quark-antiquark 
discrimination while the direction of the nearest jet to the lepton 
approximates the direction of the underlying quark.

Electrons are identified with both calorimeter and CRID information,
and this information is incorporated in a neural network trained on 
Monte Carlo electrons. Muon identification uses information from tracking,
the Warm Iron Calorimeter, and the CRID.
Both electron and muon identification algorithms have been tested 
on control samples from data.
If an electron is identified in an event, a secondary vertex is
required in either hemisphere to reject $uds$ events.
If a muon is identified, the vertex mass and the $L/D$ variable,
illustrated in Figure~\ref{LT_zvtop}, are included in the probability
function together with the total and transverse lepton momenta.
Figure~\ref{LT_Acb_lep968} shows how $L/D$ distinguishes between 
muons from direct and
cascade $b$ decays. The topological vertexing algorithm often 
finds only one vertex for both the $B$ and $D$ decay, thus 
$L/D < 1$ ($> 1$) indicates a muon from direct (cascade) $b$ decay.
$A_b$ and $A_c$ are determined simultaneously from a maximum likelihood
analysis. We measure $A_b = 0.924 \pm 0.032(stat.) \pm 0.026(syst.)$
and $A_c = 0.567 \pm 0.051(stat.) \pm 0.064(syst.)$.

 \begin{figure}[htb]
 \begin{minipage}[b]{7cm}
 \centering
 \epsfysize=5cm
 \leavevmode
 \epsfbox{LT_zvtop.epsi}
 \caption{\label{LT_zvtop}
          Topological track parameters:
          $D$ is the distance of the secondary vertex seed from the IP, 
          $L$ is the distance from the IP
          of the projection of the track's point of closest approach 
          on the vertex axis.}
 \end{minipage}
 \hfill
 \begin{minipage}[b]{7cm}
 \centering
 \epsfysize=7cm
 \leavevmode
 \epsfbox{LT_Acb_lep968.epsi}
 \caption{\label{LT_Acb_lep968}
          Tails of the $L/D$ distribution for muons in data (dots)
          and Monte Carlo (histograms).}
 \end{minipage}
 \end{figure}

\subsection{$A_c$ with Exclusive $D$ mesons}
This analysis~\cite{Ac_exDslpi} exlusively reconstructs 
six modes to tag the charm quark:
$D^+ \rightarrow K^- \pi^+ \pi^+$, $D^0 \rightarrow K^- \pi^+$,
and $D^{*+} \rightarrow D^0 \pi^+_{soft}$ with $D^0$ decaying 
into $K^- \pi^+$, $K^- \pi^+ \pi^0$, $K^- \pi^+ \pi^+ \pi^- $,
$K^- l^+ \nu_l$ ($l = e$ or $\mu$). Both $b$ and $uds$ backgrounds 
are rejected with vertex information.
The reconstruction efficiency is $\sim$4\%, however, the high 
analyzing power and the good determination of the underlying 
charm quark direction lead to low systematic errors.
Figure~\ref{LT_Ac_exclDst968} shows the distribution of the 
mass difference between $D^{*+}$ and $D^0$. The background under the
signal is estimated from the sidebands. 
We obtain $A_c = 0.690 \pm 0.042(stat.) \pm 0.022(syst.)$.

\subsection{$A_c$ with Inclusive Soft Pion}
In this analysis~\cite{Ac_exDslpi} the charm quark is tagged by the presence
of a slow pion from the $D^{*+} \rightarrow D^0 \pi^+_{soft}$
decay. The soft pion in this decay is produced along the
$D^{*+}$ jet direction ($P^2_T \sim 0$).
Figure~\ref{LT_Ac_slpi968} illustrates the $P^2_T$ distribution 
for soft pion tracks. A signal to background ratio of $1:2$ is achieved for
$P^2_T < 0.01$~(GeV/c)$^2$. This method yields 
$A_c = 0.683 \pm 0.052(stat.) \pm 0.050(syst.)$.
 \begin{figure}[p]
 \centering
 \epsfxsize8cm
 \leavevmode
 \epsffile{LT_Ac_exclDst968.epsi}
 \caption {\label{LT_Ac_exclDst968}
          The mass difference distributions for the decay of
          $D^{*+} \rightarrow D^0 \pi^+_{soft}$ with 
          $D^0$ decaying into $K^- \pi^+$, $K^- \pi^+ \pi^0$, 
          $K^- \pi^+ \pi^+ \pi^- $, $K^- l^+ \nu_l$ ($l = e$ or $\mu$) in 
          the data (dots) and Monte Carlo (histograms).}
 \vspace{1.5cm}
 \centering
 \epsfxsize8cm
 \leavevmode
 \epsffile{LT_Ac_slpi968.epsi}
 \caption {\label{LT_Ac_slpi968}
           $P^2_T$ distribution of the soft pion candidates
           with (left) the full sample and (right) after 
           background subtraction. Dots represent the data.}
 \end{figure}

\subsection{$A_c$ with Vertex Charge and Identified Kaons}
\begin{figure}[ht]
 \centering
 \epsfxsize7cm
 \leavevmode
 \epsffile{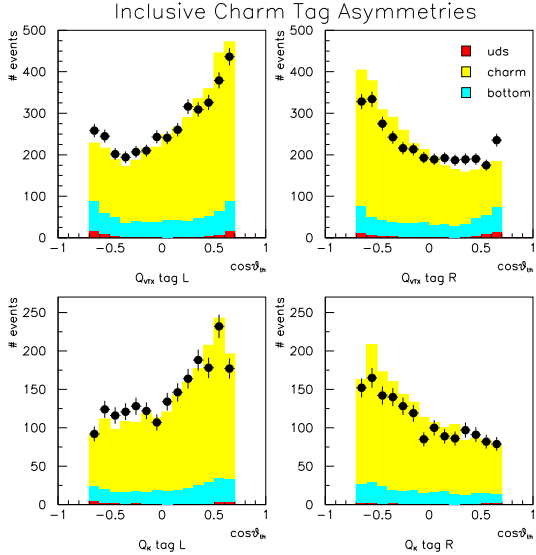}
 \caption {\label{LT_Ac_incl938}
           Polar angle distributions of the signed thrust axis
           for (top) the vertex charge and (bottom) kaon charge 
           channels for data (dots) and Monte Carlo (histograms).
           The left and right sides show the distributions for
           the left-handed and right-handed electron beams, respectively.}
 \end{figure}

This analysis~\cite{Ac_vtxKtag} uses the charm tag described for the $R_c$ 
analysis. Furthermore we require at least one hemisphere
to pass the charm selection while neither hemisphere
passes the bottom selection. 
The sign of the quark charge is determined by the charge
of an identified $K^{\pm}$ (or by vertex charge), present 
in $\sim$25\% ($\sim$50\%) of the selected events with
more than 90\% correct sign fraction.
Figure~\ref{LT_Ac_incl938} shows the polar angle distributions 
of the signed thrust axis for left-handed and right-handed electron beams. 
The analyzing power for $c$ events is calibrated from the data.
We obtain $A_c = 0.603 \pm 0.028(stat.) \pm 0.023(syst.)$.

\subsection{$A_s$ Measurement}
\begin{figure}[htb]
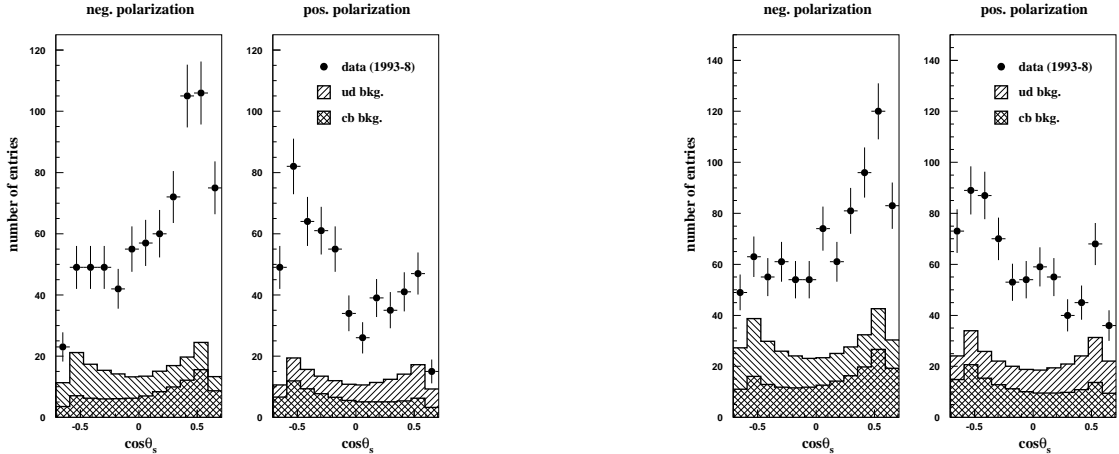

 \begin{minipage}[b]{7cm}
 \centering
 \epsfysize=6cm
 \leavevmode
 \epsfbox{LT_As_KpKm938.epsi}
 \end{minipage}
 \hfill
 \begin{minipage}[b]{7cm}
 \centering
 \epsfysize=6cm
 \leavevmode
 \epsfbox{LT_As_KpmKs938.epsi}
 \end{minipage}
 \caption{\label{LT_As_KpKm_KpmKs}
           Polar angle distributions of the signed thrust axis
           for the (left) $K^+ K^-$ mode and (right) $K^{\pm} K^0_s$  
           mode for left-handed and right-handed electron beams.
           The dots show data and the background is indicated
           by the hatched histograms.}
 \end{figure}

At SLD, $s\bar{s}$ events can be identified with relatively
high purity due to the good separation of tracks from secondary 
vertices and the CRID particle identification.
In the $A_s$ analysis~\cite{As}, the charge of an identified 
$K^{\pm}$ is used to tag the sign of the initial $s$ quark.
Information from the vertex detector is used to suppress
the background from heavy flavor events.
$K^{\pm}$ with $p > 9$~GeV/c and $K^0_s$ with $p > 5$~GeV/c
are selected with 92\% and 91\% purity, respectively.
Each thrust hemisphere of a light flavor tagged event 
is required to contain at least one identified strange 
particle. Strange hemispheres are tagged using the highest
momentum strange particle present in the hemisphere.
A tag is required in both hemispheres and at least one tag 
must be signed; if both are signed, signs must be opposite.
The combined $s\bar{s}$ purity of the $K^+ K^-$ and $K^{\pm} K^0_s$ 
tagging modes is 66\%.
The initial $s$ quark direction is approximated by the 
thrust axis in the event, signed to point in the direction
of negative strangeness. 
Figure~\ref{LT_As_KpKm_KpmKs} shows the polar angle distributions
for left-handed and right-handed electron beams.
The background from $ud$
events as well as the analyzing power of the method for 
$s$ events are constrained from the  data.
We measure $A_s = 0.895 \pm 0.066(stat.) \pm 0.063(syst.)$.

\section{Conclusions}
We have presented the results of several electroweak
measurements performed by the SLD Collaboration.
These results are summarized in Table~\ref{SLDresults}.
The combined $A_{LR}$ and $A_{lepton}$ SLD measurements yield 
$\sin^2{\theta_W^{eff}} = 0.23099 \pm 0.00026$. 
The LEP measurements of lepton forward-backward
asymmetries and tau polarization have been combined
into a LEP lepton based 
$\sin^2{\theta_W^{eff}} = 0.23151 \pm 0.00033$. 
These results are consistent.
Both the $R_c$ and $R_b$ SLD results are in good agreement with
the LEP results and the Standard Model prediction.
The SLD $R_c$ measurement is the most precise single
determination of this variable. 
\begin{table}[ht]
\begin{center}
\caption{\label{SLDresults}
         Summary of SLD results of electroweak measurements.}
\bigskip
\begin{tabular}{|c||c|} 
\hline
 Observable & Preliminary Result \\
\hline \hline
 $A_{LR}^0$                                  & $0.15108 \pm 0.00218$  \\
 $\sin^2{\theta_W^{eff}}$ (from $A_{LR}^0$)  & $0.23101 \pm 0.00028$  \\
\hline
 $A_e$ (from $\tilde{A}^f_{FB}$)             & $0.1558 \pm 0.0064$    \\
 $A_{\mu}$                                   & $0.137 \pm 0.016$      \\
 $A_{\tau}$                                  & $0.142 \pm 0.016$      \\
 $A_{e\mu\tau}$                              & $0.1523 \pm 0.0057$    \\
 $\sin^2{\theta_W^{eff}}$ (from $A_l$)       & $0.23085 \pm 0.00073$  \\
\hline
 $\sin^2{\theta_W^{eff}}$ (from $A_{LR}^0$ and $A_l$) 
                                             & $0.23099 \pm 0.00026$  \\
\hline
 $R_c$                                       & $0.1685 \pm 0.0047 \pm 0.0043$ \\
 $R_b$                                       & $0.2159 \pm 0.0014 \pm 0.0014$ \\
\hline
 $A_b$ (kaon tag)                            & $0.960 \pm 0.040 \pm 0.069$  \\
 $A_b$ (vertex charge)                       & $0.897 \pm 0.027 \pm 0.034$  \\
 $A_b$ (jet charge)                          & $0.882 \pm 0.020 \pm 0.029$  \\
 $A_b$ (lepton tag)                          & $0.924 \pm 0.032 \pm 0.026$  \\
\hline
 $A_b$ (SLD average)                         & $0.905 \pm 0.026$            \\
\hline
 $A_c$ (lepton tag)                          & $0.567 \pm 0.051 \pm 0.064$  \\
 $A_c$ (exclusive $D$, $D^*$)                & $0.690 \pm 0.042 \pm 0.022$  \\ 
 $A_c$ (inclusive soft pion)                 & $0.683 \pm 0.052 \pm 0.050$  \\
 $A_c$ (vertex charge, kaons)                & $0.603 \pm 0.028 \pm 0.023$  \\
\hline
 $A_c$ (SLD average)                         & $0.634 \pm 0.027$            \\ 
\hline
 $A_s$                                       & $0.895 \pm 0.066 \pm 0.063$  \\ 
\hline
\end{tabular}
\end{center}
\end{table}

The SLD $A_s$ result is in agreement with LEP results
and the Standard model prediction. It is also consistent
with previous $A_b$ measurements performed by SLD and LEP,
and therefore supports the predicted universality of the
$Z^0$ to down-type quark couplings.
The combined SLD $A_c$ measurements give $A_c = 0.634 \pm 0.027$,
in agreement with both the corresponding LEP result and 
the Standard Model prediction.
The combined SLD $A_b$ results yield $A_b = 0.905 \pm 0.026$
which is consistent with the Standard Model prediction.
The corresponding LEP result is $A_b = 0.881 \pm 0.020$.
A combined SLD and LEP average for $A_b$ is about 2.8 standard 
deviations below the Standard Model prediction.

\section{Acknowledgements}
I thank all SLD collaborators for their support and efforts
and the SLAC accelerator department for its outstanding performance.
I thank the organizers of Les Rencontres de la Vallee d'Aoste
for inviting me and for the excellent hospitality.
This work was supported in part by DOE Contract DE-AC03-76SF00515~(SLAC).


\section*{$^{**}$List of Authors}
%
%
%
\begin{center}
\def\iADEL{$^{(1)}$}
\def\iAOMORI{$^{(2)}$}
\def\iBOLO{$^{(3)}$}
\def\iBRI{$^{(4)}$}
\def\iBRUN{$^{(5)}$}
\def\iBU{$^{(6)}$}
\def\iCINC{$^{(7)}$}
\def\iCOLO{$^{(8)}$}
\def\iCOLU{$^{(9)}$}
\def\iCSU{$^{(10)}$}
\def\iFERR{$^{(11)}$}
\def\iFRAS{$^{(12)}$}
\def\iILLI{$^{(13)}$}
\def\iJHU{$^{(14)}$}
\def\iLBL{$^{(15)}$}
\def\iLTU{$^{(16)}$}
\def\iMASS{$^{(17)}$}
\def\iMISSI{$^{(18)}$}
\def\iMIT{$^{(19)}$}
\def\iMOSCOW{$^{(20)}$}
\def\iNAGO{$^{(21)}$}
\def\iOREG{$^{(22)}$}
\def\iOXF{$^{(23)}$}
\def\iPADO{$^{(24)}$}
\def\iPERU{$^{(25)}$}
\def\iPISA{$^{(26)}$}
\def\iRAL{$^{(27)}$}
\def\iRUTG{$^{(28)}$}
\def\iSLAC{$^{(29)}$}
\def\iSOGA{$^{(30)}$}
\def\iSOONG{$^{(31)}$}
\def\iTENN{$^{(32)}$}
\def\iTOHO{$^{(33)}$}
\def\iUCSB{$^{(34)}$}
\def\iUCSC{$^{(35)}$}
\def\iUVIC{$^{(36)}$}
\def\iVAND{$^{(37)}$}
\def\iWASH{$^{(38)}$}
\def\iWISC{$^{(39)}$}
\def\iYALE{$^{(40)}$}

  \baselineskip=.75\baselineskip  
\mbox{Kenji  Abe\unskip,\iNAGO}
\mbox{Koya Abe\unskip,\iTOHO}
\mbox{T. Abe\unskip,\iSLAC}
\mbox{I.Adam\unskip,\iSLAC}
\mbox{T.  Akagi\unskip,\iSLAC}
\mbox{N. J. Allen\unskip,\iBRUN}
\mbox{W.W. Ash\unskip,\iSLAC}
\mbox{D. Aston\unskip,\iSLAC}
\mbox{K.G. Baird\unskip,\iMASS}
\mbox{C. Baltay\unskip,\iYALE}
\mbox{H.R. Band\unskip,\iWISC}
\mbox{M.B. Barakat\unskip,\iLTU}
\mbox{O. Bardon\unskip,\iMIT}
\mbox{T.L. Barklow\unskip,\iSLAC}
\mbox{G. L. Bashindzhagyan\unskip,\iMOSCOW}
\mbox{J.M. Bauer\unskip,\iMISSI}
\mbox{G. Bellodi\unskip,\iOXF}
\mbox{R. Ben-David\unskip,\iYALE}
\mbox{A.C. Benvenuti\unskip,\iBOLO}
\mbox{G.M. Bilei\unskip,\iPERU}
\mbox{D. Bisello\unskip,\iPADO}
\mbox{G. Blaylock\unskip,\iMASS}
\mbox{J.R. Bogart\unskip,\iSLAC}
\mbox{G.R. Bower\unskip,\iSLAC}
\mbox{J. E. Brau\unskip,\iOREG}
\mbox{M. Breidenbach\unskip,\iSLAC}
\mbox{W.M. Bugg\unskip,\iTENN}
\mbox{D. Burke\unskip,\iSLAC}
\mbox{T.H. Burnett\unskip,\iWASH}
\mbox{P.N. Burrows\unskip,\iOXF}
\mbox{A. Calcaterra\unskip,\iFRAS}
\mbox{D. Calloway\unskip,\iSLAC}
\mbox{B. Camanzi\unskip,\iFERR}
\mbox{M. Carpinelli\unskip,\iPISA}
\mbox{R. Cassell\unskip,\iSLAC}
\mbox{R. Castaldi\unskip,\iPISA}
\mbox{A. Castro\unskip,\iPADO}
\mbox{M. Cavalli-Sforza\unskip,\iUCSC}
\mbox{A. Chou\unskip,\iSLAC}
\mbox{E. Church\unskip,\iWASH}
\mbox{H.O. Cohn\unskip,\iTENN}
\mbox{J.A. Coller\unskip,\iBU}
\mbox{M.R. Convery\unskip,\iSLAC}
\mbox{V. Cook\unskip,\iWASH}
\mbox{R. Cotton\unskip,\iBRUN}
\mbox{R.F. Cowan\unskip,\iMIT}
\mbox{D.G. Coyne\unskip,\iUCSC}
\mbox{G. Crawford\unskip,\iSLAC}
\mbox{C.J.S. Damerell\unskip,\iRAL}
\mbox{M. N. Danielson\unskip,\iCOLO}
\mbox{M. Daoudi\unskip,\iSLAC}
\mbox{N. de Groot\unskip,\iBRI}
\mbox{R. Dell'Orso\unskip,\iPERU}
\mbox{P.J. Dervan\unskip,\iBRUN}
\mbox{R. de Sangro\unskip,\iFRAS}
\mbox{M. Dima\unskip,\iCSU}
\mbox{A. D'Oliveira\unskip,\iCINC}
\mbox{D.N. Dong\unskip,\iMIT}
\mbox{M. Doser\unskip,\iSLAC}
\mbox{R. Dubois\unskip,\iSLAC}
\mbox{B.I. Eisenstein\unskip,\iILLI}
\mbox{V. Eschenburg\unskip,\iMISSI}
\mbox{E. Etzion\unskip,\iWISC}
\mbox{S. Fahey\unskip,\iCOLO}
\mbox{D. Falciai\unskip,\iFRAS}
\mbox{C. Fan\unskip,\iCOLO}
\mbox{J.P. Fernandez\unskip,\iUCSC}
\mbox{M.J. Fero\unskip,\iMIT}
\mbox{K.Flood\unskip,\iMASS}
\mbox{R. Frey\unskip,\iOREG}
\mbox{J. Gifford\unskip,\iUVIC}
\mbox{T. Gillman\unskip,\iRAL}
\mbox{G. Gladding\unskip,\iILLI}
\mbox{S. Gonzalez\unskip,\iMIT}
\mbox{E. R. Goodman\unskip,\iCOLO}
\mbox{E.L. Hart\unskip,\iTENN}
\mbox{J.L. Harton\unskip,\iCSU}
\mbox{A. Hasan\unskip,\iBRUN}
\mbox{K. Hasuko\unskip,\iTOHO}
\mbox{S. J. Hedges\unskip,\iBU}
\mbox{S.S. Hertzbach\unskip,\iMASS}
\mbox{M.D. Hildreth\unskip,\iSLAC}
\mbox{J. Huber\unskip,\iOREG}
\mbox{M.E. Huffer\unskip,\iSLAC}
\mbox{E.W. Hughes\unskip,\iSLAC}
\mbox{X.Huynh\unskip,\iSLAC}
\mbox{H. Hwang\unskip,\iOREG}
\mbox{M. Iwasaki\unskip,\iOREG}
\mbox{D. J. Jackson\unskip,\iRAL}
\mbox{P. Jacques\unskip,\iRUTG}
\mbox{J.A. Jaros\unskip,\iSLAC}
\mbox{Z.Y. Jiang\unskip,\iSLAC}
\mbox{A.S. Johnson\unskip,\iSLAC}
\mbox{J.R. Johnson\unskip,\iWISC}
\mbox{R.A. Johnson\unskip,\iCINC}
\mbox{T. Junk\unskip,\iSLAC}
\mbox{R. Kajikawa\unskip,\iNAGO}
\mbox{M. Kalelkar\unskip,\iRUTG}
\mbox{Y. Kamyshkov\unskip,\iTENN}
\mbox{H.J. Kang\unskip,\iRUTG}
\mbox{I. Karliner\unskip,\iILLI}
\mbox{H. Kawahara\unskip,\iSLAC}
\mbox{Y. D. Kim\unskip,\iSOGA}
\mbox{M.E. King\unskip,\iSLAC}
\mbox{R. King\unskip,\iSLAC}
\mbox{R.R. Kofler\unskip,\iMASS}
\mbox{N.M. Krishna\unskip,\iCOLO}
\mbox{R.S. Kroeger\unskip,\iMISSI}
\mbox{M. Langston\unskip,\iOREG}
\mbox{A. Lath\unskip,\iMIT}
\mbox{D.W.G. Leith\unskip,\iSLAC}
\mbox{V. Lia\unskip,\iMIT}
\mbox{C.Lin\unskip,\iMASS}
\mbox{M.X. Liu\unskip,\iYALE}
\mbox{X. Liu\unskip,\iUCSC}
\mbox{M. Loreti\unskip,\iPADO}
\mbox{A. Lu\unskip,\iUCSB}
\mbox{H.L. Lynch\unskip,\iSLAC}
\mbox{J. Ma\unskip,\iWASH}
\mbox{G. Mancinelli\unskip,\iRUTG}
\mbox{S. Manly\unskip,\iYALE}
\mbox{G. Mantovani\unskip,\iPERU}
\mbox{T.W. Markiewicz\unskip,\iSLAC}
\mbox{T. Maruyama\unskip,\iSLAC}
\mbox{H. Masuda\unskip,\iSLAC}
\mbox{E. Mazzucato\unskip,\iFERR}
\mbox{A.K. McKemey\unskip,\iBRUN}
\mbox{B.T. Meadows\unskip,\iCINC}
\mbox{G. Menegatti\unskip,\iFERR}
\mbox{R. Messner\unskip,\iSLAC}
\mbox{P.M. Mockett\unskip,\iWASH}
\mbox{K.C. Moffeit\unskip,\iSLAC}
\mbox{T.B. Moore\unskip,\iYALE}
\mbox{M.Morii\unskip,\iSLAC}
\mbox{D. Muller\unskip,\iSLAC}
\mbox{V.Murzin\unskip,\iMOSCOW}
\mbox{T. Nagamine\unskip,\iTOHO}
\mbox{S. Narita\unskip,\iTOHO}
\mbox{U. Nauenberg\unskip,\iCOLO}
\mbox{H. Neal\unskip,\iSLAC}
\mbox{M. Nussbaum\unskip,\iCINC}
\mbox{N.Oishi\unskip,\iNAGO}
\mbox{D. Onoprienko\unskip,\iTENN}
\mbox{L.S. Osborne\unskip,\iMIT}
\mbox{R.S. Panvini\unskip,\iVAND}
\mbox{C. H. Park\unskip,\iSOONG}
\mbox{T.J. Pavel\unskip,\iSLAC}
\mbox{I. Peruzzi\unskip,\iFRAS}
\mbox{M. Piccolo\unskip,\iFRAS}
\mbox{L. Piemontese\unskip,\iFERR}
\mbox{K.T. Pitts\unskip,\iOREG}
\mbox{R.J. Plano\unskip,\iRUTG}
\mbox{R. Prepost\unskip,\iWISC}
\mbox{C.Y. Prescott\unskip,\iSLAC}
\mbox{G.D. Punkar\unskip,\iSLAC}
\mbox{J. Quigley\unskip,\iMIT}
\mbox{B.N. Ratcliff\unskip,\iSLAC}
\mbox{T.W. Reeves\unskip,\iVAND}
\mbox{J. Reidy\unskip,\iMISSI}
\mbox{P.L. Reinertsen\unskip,\iUCSC}
\mbox{P.E. Rensing\unskip,\iSLAC}
\mbox{L.S. Rochester\unskip,\iSLAC}
\mbox{P.C. Rowson\unskip,\iCOLU}
\mbox{J.J. Russell\unskip,\iSLAC}
\mbox{O.H. Saxton\unskip,\iSLAC}
\mbox{T. Schalk\unskip,\iUCSC}
\mbox{R.H. Schindler\unskip,\iSLAC}
\mbox{B.A. Schumm\unskip,\iUCSC}
\mbox{J. Schwiening\unskip,\iSLAC}
\mbox{S. Sen\unskip,\iYALE}
\mbox{V.V. Serbo\unskip,\iSLAC}
\mbox{M.H. Shaevitz\unskip,\iCOLU}
\mbox{J.T. Shank\unskip,\iBU}
\mbox{G. Shapiro\unskip,\iLBL}
\mbox{D.J. Sherden\unskip,\iSLAC}
\mbox{K. D. Shmakov\unskip,\iTENN}
\mbox{C. Simopoulos\unskip,\iSLAC}
\mbox{N.B. Sinev\unskip,\iOREG}
\mbox{S.R. Smith\unskip,\iSLAC}
\mbox{M. B. Smy\unskip,\iCSU}
\mbox{J.A. Snyder\unskip,\iYALE}
\mbox{H. Staengle\unskip,\iCSU}
\mbox{A. Stahl\unskip,\iSLAC}
\mbox{P. Stamer\unskip,\iRUTG}
\mbox{H. Steiner\unskip,\iLBL}
\mbox{R. Steiner\unskip,\iADEL}
\mbox{M.G. Strauss\unskip,\iMASS}
\mbox{D. Su\unskip,\iSLAC}
\mbox{F. Suekane\unskip,\iTOHO}
\mbox{A. Sugiyama\unskip,\iNAGO}
\mbox{S. Suzuki\unskip,\iNAGO}
\mbox{M. Swartz\unskip,\iJHU}
\mbox{A. Szumilo\unskip,\iWASH}
\mbox{T. Takahashi\unskip,\iSLAC}
\mbox{F.E. Taylor\unskip,\iMIT}
\mbox{J. Thom\unskip,\iSLAC}
\mbox{E. Torrence\unskip,\iMIT}
\mbox{N. K. Toumbas\unskip,\iSLAC}
\mbox{T. Usher\unskip,\iSLAC}
\mbox{C. Vannini\unskip,\iPISA}
\mbox{J. Va'vra\unskip,\iSLAC}
\mbox{E. Vella\unskip,\iSLAC}
\mbox{J.P. Venuti\unskip,\iVAND}
\mbox{R. Verdier\unskip,\iMIT}
\mbox{P.G. Verdini\unskip,\iPISA}
\mbox{D. L. Wagner\unskip,\iCOLO}
\mbox{S.R. Wagner\unskip,\iSLAC}
\mbox{A.P. Waite\unskip,\iSLAC}
\mbox{S. Walston\unskip,\iOREG}
\mbox{J.Wang\unskip,\iSLAC}
\mbox{S.J. Watts\unskip,\iBRUN}
\mbox{A.W. Weidemann\unskip,\iTENN}
\mbox{E. R. Weiss\unskip,\iWASH}
\mbox{J.S. Whitaker\unskip,\iBU}
\mbox{S.L. White\unskip,\iTENN}
\mbox{F.J. Wickens\unskip,\iRAL}
\mbox{B. Williams\unskip,\iCOLO}
\mbox{D.C. Williams\unskip,\iMIT}
\mbox{S.H. Williams\unskip,\iSLAC}
\mbox{S. Willocq\unskip,\iMASS}
\mbox{R.J. Wilson\unskip,\iCSU}
\mbox{W.J. Wisniewski\unskip,\iSLAC}
\mbox{J. L. Wittlin\unskip,\iMASS}
\mbox{M. Woods\unskip,\iSLAC}
\mbox{G.B. Word\unskip,\iVAND}
\mbox{T.R. Wright\unskip,\iWISC}
\mbox{J. Wyss\unskip,\iPADO}
\mbox{R.K. Yamamoto\unskip,\iMIT}
\mbox{J.M. Yamartino\unskip,\iMIT}
\mbox{X. Yang\unskip,\iOREG}
\mbox{J. Yashima\unskip,\iTOHO}
\mbox{S.J. Yellin\unskip,\iUCSB}
\mbox{C.C. Young\unskip,\iSLAC}
\mbox{H. Yuta\unskip,\iAOMORI}
\mbox{G. Zapalac\unskip,\iWISC}
\mbox{R.W. Zdarko\unskip,\iSLAC}
\mbox{J. Zhou\unskip.\iOREG}

\it
  \vskip \baselineskip                   
  \vskip \baselineskip        
  \baselineskip=.75\baselineskip   
\iADEL
  Adelphi University, Garden City, New York 11530, \break
\iAOMORI
  Aomori University, Aomori , 030 Japan, \break
\iBOLO
  INFN Sezione di Bologna, I-40126, Bologna Italy, \break
\iBRI
  University of Bristol, Bristol, U.K., \break
\iBRUN
  Brunel University, Uxbridge, Middlesex, UB8 3PH United Kingdom, \break
\iBU
  Boston University, Boston, Massachusetts 02215, \break
\iCINC
  University of Cincinnati, Cincinnati, Ohio 45221, \break
\iCOLO
  University of Colorado, Boulder, Colorado 80309, \break
\iCOLU
  Columbia University, New York, New York 10533, \break
\iCSU
  Colorado State University, Ft. Collins, Colorado 80523, \break
\iFERR
  INFN Sezione di Ferrara and Universita di Ferrara, I-44100 Ferrara, Italy, \break
\iFRAS
  INFN Lab. Nazionali di Frascati, I-00044 Frascati, Italy, \break
\iILLI
  University of Illinois, Urbana, Illinois 61801, \break
\iJHU
  Johns Hopkins University, Baltimore, MD 21218-2686, \break
\iLBL
  Lawrence Berkeley Laboratory, University of California, Berkeley, California 94720, \break
\iLTU
  Louisiana Technical University - Ruston,LA 71272, \break
\iMASS
  University of Massachusetts, Amherst, Massachusetts 01003, \break
\iMISSI
  University of Mississippi, University, Mississippi 38677, \break
\iMIT
  Massachusetts Institute of Technology, Cambridge, Massachusetts 02139, \break
\iMOSCOW
  Institute of Nuclear Physics, Moscow State University, 119899, Moscow Russia, \break
\iNAGO
  Nagoya University, Chikusa-ku, Nagoya 464 Japan, \break
\iOREG
  University of Oregon, Eugene, Oregon 97403, \break
\iOXF
  Oxford University, Oxford, OX1 3RH, United Kingdom, \break
\iPADO
  INFN Sezione di Padova and Universita di Padova I-35100, Padova, Italy, \break
\iPERU
  INFN Sezione di Perugia and Universita di Perugia, I-06100 Perugia, Italy, \break
\iPISA
  INFN Sezione di Pisa and Universita di Pisa, I-56010 Pisa, Italy, \break
\iRAL
  Rutherford Appleton Laboratory, Chilton, Didcot, Oxon OX11 0QX United Kingdom, \break
\iRUTG
  Rutgers University, Piscataway, New Jersey 08855, \break
\iSLAC
  Stanford Linear Accelerator Center, Stanford University, Stanford, California 94309, \break
\iSOGA
  Sogang University, Seoul, Korea, \break
\iSOONG
  Soongsil University, Seoul, Korea 156-743, \break
\iTENN
  University of Tennessee, Knoxville, Tennessee 37996, \break
\iTOHO
  Tohoku University, Sendai 980, Japan, \break
\iUCSB
  University of California at Santa Barbara, Santa Barbara, California 93106, \break
\iUCSC
  University of California at Santa Cruz, Santa Cruz, California 95064, \break
\iUVIC
  University of Victoria, Victoria, B.C., Canada, V8W 3P6, \break
\iVAND
  Vanderbilt University, Nashville,Tennessee 37235, \break
\iWASH
  University of Washington, Seattle, Washington 98105, \break
\iWISC
  University of Wisconsin, Madison,Wisconsin 53706, \break
\iYALE
  Yale University, New Haven, Connecticut 06511. \break

\rm
%

\end{center}


\end{document}